\documentstyle[12pt]{article}
\input{tcilatex}
\begin{document}

\noindent {\Large {\bf Low energy Ar$^{+}$ ion beam induced kinetic
roughening of thin Pt films on a Si substrate}} \vspace{0.7in}\newline

\begin{center}
{\bf P. Karmakar and D. Ghose\footnote{{\bf Corresponding author. e-mail:
ghose@surf.saha.ernet.in}} } \vspace{0.1in} \\[0pt]
{\it Saha Institute of Nuclear Physics,\\[0pt]
Sector - I, Block - AF, Bidhan Nagar, Kolkata 700064, India} \vspace{0.4in}\\%
[0pt]
\today \vspace{0.2in}\\[0pt]
{\bf Abstract}\\[0pt]
\end{center}

A 30 $nm$ Pt thin film evaporated onto a Si wafer was sputtered by 8 keV Ar$%
^{+}$ ions at various ion doses. The evolution of the modified sputtered
films was monitored by atomic force microscopy (AFM), high resolution
scanning electron microscopy (HRSEM) and Rutherford backscattering
spectrometry (RBS). The most interesting observation was the formation of
mound-like structures on the metal surface. Morphological data were
quantitatively analysed within the framework of the dynamic scaling theory.
Analyses of the height-height correlation function for different doses yield
roughness exponents $\alpha $ in the range 0.65 - 0.87, while the
root-mean-square roughness amplitude $w$ evolves with the dose $\phi $ as a
power law $w\propto \phi ^{\beta }$, with the growth exponent, $\beta $ $%
\approx $ $0.3$. The results are discussed.\newline

PACS numbers: 68.35.Ct, 79.20.Rf, 61.80.Jh

\pagebreak

\begin{center}
I. INTRODUCTION
\end{center}

{\noindent} Low energy ion beam (typically 1 - 10 keV) sputtering is
commonly used in most of the analytical depth-profiling techniques including
secondary ion mass spectrometry (SIMS) and Auger electron spectroscopy
(AES). Also, low energy ion beam system forms an essential part in various
surface processing techniques such as ion beam etching, deposition and ion
beam assisted growth. During the bombardment process, the interaction
between the projectile ion and the target atoms lead to different phenomena,
e.g., the topographical modification of the sample surface. A large variety
of characteristic surface structures ranging from stochastic rough surface
to the formation of well-defined conical protrusions [1, 2] are found to
develop on the sputtered surfaces. Recently, there is a growing interest for
the fabrication of sputtering induced surface structures in the submicron
length scale because of possible technological applications [3]. Different
scaling theories for the evolution of the surface roughness have been
proposed [4] in order to understand the underlying physical processes
involved and eventually determining a way of optimising control of the
process.\newline

In this paper, we have investigated the development of ion beam
sputtering-induced surface morphologies in thin Pt films evaporated onto Si
substrates. The analyses of the irradiated samples were done by atomic force
microscopy (AFM), high resolution scanning electron microscopy (HRSEM) and
Rutherford backscattering spectrometry (RBS).\newline

\begin{center}
\smallskip

II. EXPERIMENTAL
\end{center}

{\noindent } Thin Pt films ($\simeq 30$ $nm$) were deposited by d. c.
magnetron sputtering onto commercially available polished Si (100) wafers,
previously degreased and cleaned. The base pressure in the deposition
chamber was $2\times 10^{-6}$ mbar. The Pt/Si samples were then sputtered
with 8 keV Ar$^{+}$ ions at different doses in a low energy ion beam set-up
developed in the laboratory [5]. The angle of ion incidence with respect to
the surface normal was $45^{0}$. The average current density was 10 ${\mu }%
A/cm^{2}$. The ion dose was measured by a current integrator (Danfysik,
model 554) after suppression of the secondary electrons. The samples were
exposed to total ion doses between $1\times 10^{15}$ and $2\times 10^{18}$ $%
ions/cm^{2}$. The base pressure in the target chamber was less than $5\times
10^{-8}$ mbar. The surface morphology of the ion-irradiated samples was
examined by a Park Scientific AFM (Auto Probe CP) as well as by a high
resolution scanning electron microscope (Hitachi S-4700) operated at 15 kV
and viewed at $20^{0}$. Compositional profiles in the samples were analyzed
by means of Rutherford backscattering of 2 MeV He$^{2+}$ ions with the
scattering angle set at $110^{0}$.\newline

\begin{center}
\smallskip

III. RESULTS AND DISCUSSION
\end{center}

{\noindent } Atomic force microscopy of the initial Pt surface topography
shows a smmoth continuous film with a root-mean-square (rms) roughness $%
w\simeq 0.2$ $nm$. Fig. 1 shows the AFM images from the sample surfaces
sputtered at successive increasing ion doses $\phi $ (different bombarding
times $t$). Immediately after the start of the bombardment mound-like or
globular structures begin to appear. The bombardment also produces voids or
vacant regions in between the mounds. Both the lateral size and the height
of the mounds become larger with increasing ion dose upto $\sim 10^{17}$ $%
ions/cm^{2}$; thereafter the sizes tend to decrease rapidly as the
sputtering further continues. In the fluence region $10^{16}$ $ions/cm^{2}$
the mounds are quite sharp and well developed. However, at doses $\ge
2\times 10^{16}$ $ions/cm^{2}$, the mounds tend to form clusters and become
blurred. In a separate experiment we have bombarded a clean Si (100) wafer
under identical conditions at various doses upto $10^{19}$ $ions/cm^{2}$.
But in no case we find development of any surface structure in the virgin Si
surface within the present AFM resolution, as also reported by Vajo et al.
[6]. Therefore, it appears that the observed morphological structures are
the characteristics of the Pt films. The sputtered surface topography was
also investigated simultaneously using HRSEM. Although the image resolution
and contrast are not as good as that of AFM, the results also show similar
topographical evolution. For comparison typical HRSEM images at two
different ion doses are shown in Fig. 2. Finally, comparison of the RBS
spectra of the as-deposited sample and the ion-irradiated samples show that
both the height and the width of the Pt profile reduce gradually due to ion
beam sputtering, while the implanted Ar peak tends to develop as the
bombardment is continued (Fig. 3). Energy dispersive x-ray analysis of the
irradiated samples carried out in HRSEM also confirms the erosion of Pt due
to Ar$^{+}$ bombardment.\newline

In order to quantitatively characterize the observed morphology, we have
studied the scaling properties of the interface. Family and Vicsek [7]
analyzed the behavior of growing surfaces by assuming that they were
self-affine such that the root-mean-square (rms) roughness or the interface
width, $w(t)$ obeys the relation: 
\begin{equation}
w(L,\text{ }t)=L^{\alpha }\text{ }f(t/L^{z}),
\end{equation}
{\noindent }where $L$ is the length scale over which the roughness is
measured and $t$ is the elapsed time of growth. The scaling function f(u)
behaves as $u^{\beta }$ for $u\ll 1$ and as a constant for $u\gg 1$. The two
parameters $\alpha $ and $\beta $ are called the static (or spatial) and
dynamic (or temporal) scaling exponents, respectively, and z is equal to $%
\alpha /\beta $. A standard method for investigating surface morphology is
to study the height-height correlation function $G({\bf r},$ $t)$, which is
the mean square of height difference between two surface positions separated
by a lateral distance ${\bf r}$: 
\begin{equation}
G({\bf r},\text{ }t)={\langle {[h({\bf r},}}\text{ }{t)-h(0,}\text{ }{{t)]}%
^{2}\rangle }.
\end{equation}
If the surface is scale invariant and isotropic, then $G({\bf r},$ $t)$ has
the following properties:

\begin{equation}
G(r,\text{ }t)\sim \left\{ \QATOP{r^{2\alpha }\text{ for }r\ll {\xi }(t)}{%
2w^{2}(t)\text{ for }r\text{ }{\gg }\text{ }{\xi }(t)}\right. ,
\end{equation}
where ${\xi }(t)$ is the lateral correlation length which scales as $t^{1/z}$%
. w(t) is given by:

\begin{equation}
w(t)=\sqrt{\langle {[h({\bf r},}\text{ }{t)-\langle h\rangle ]}^{2}\rangle 
\text{ }}{\propto }\text{ }t^{\beta },
\end{equation}
where $\langle ..\rangle $ denotes the spatial average over the sample
surface. The width function provides a measure of the vertical height
fluctuation of the surface profile at different bombardment times $t$ $%
(\propto \phi )$. \newline

In the present case the correlation function (Eq. (2)) was evaluated
directly from the AFM micrographs by taking every possible pair of
positions, calculating the square of the height difference and averaging for
equal distances. Shown in Fig. 4 are the typical log-log plot of $G(r,$ $t)$
vs lateral distace r at three different bombarding ion doses, $\phi $. From
the figure it is seen that $G(r,$ $t)$ increases linearly at small $r$
following a plateau at large $r$, consistent with the asymptotic behavior
predicted in Eq. (3). The lateral position corresponding to the plateau
point is equal to $\xi $, which is a representative measure of the average
dimension of the mound. The roughness exponent $\alpha $ was determined by
least-squares fitting to the linear slope of $G(r,$ $t)$ at small $r$. In
the dose range investigated here we observed $\alpha $ in the range 0.65 -
0.87 (Fig. 5). The rms roughness or the interface width $w$ at each time (or
dose) was obtained from the in-built software of the AFM instrument and it
was found that they all agree to those calculated from the asymptotic value
of G(r, t) for $r$ ${\gg }$ ${\xi }(t)$ as stated in Eq. (3). These data of $%
w$ versus ion dose in a log-log plot are shown in Fig. 6. It is interesting
to note that after a certain bombarding dose when the film material is
eroded enough away, the rms surface roughness values show a sharp fall. The
exponent $\beta $ was determined from the linear fit of the ascending part
of the roughness curve and is found to be about $0.3$. Fig. 7 shows a power
law increment of the correlation length ${\xi }$ with increasing ion dose $%
\phi $, as expected, with the dynamic exponent $1/z\approx 0.39$.

The evolution of surface morphology is thought to govern by the interplay
and competition between the dynamics of surface roughening on the one hand
and material transport during surface diffusion on the other. Various
stochastic nonlinear continuum models have been introduced to explain the
temporal and spatial evolution of surface morphologies [8]. One such
formalism is the noisy Kuramoto-Sivashinsky (KS) equation [9] which is
believed to encompass most of the roughening and smoothing processes that
are occurred during ion beam sputtering of surfaces. The KS equation
describes the temporal evolution of surface height function $h$ as

\begin{equation}
\frac{\partial h}{\partial t}=\nu \nabla ^{2}h-\kappa \nabla ^{4}h+\frac{%
\lambda }{2}\left| \nabla h\right| ^{2}+\eta ,
\end{equation}

\noindent where $\nu $ is the ''negative surface tension'' generated by the
erosion process, $\kappa $ is the coefficient of thermal or ion induced
surface diffusion, $\lambda $ is a nonlinear coefficient attributed to the
tilt-dependent erosion rate, and $\eta $ is the noise term which accounts
the randomness in the arrival of the bombarding ions. This equation was
solved numerically in 2+1 dimensions [9], which yields the scaling exponents 
$\alpha $ = 0.75 - 0.8, $\beta $ = 0.22 - 0.25 and $z$ = 3.0 - 4.0,
respectively. The exponents which are observed in the present experiment
nearly correspond to the above predicted values. It is interesting to
mention here that the present results are also consistent to the scaling
laws obtained in the experiment of sputter-deposition growth of Pt by
Jeffries et al. [10]. This supports the assumption often made that the
sputter erosion is equivalent to the inverse of the growth process [4]. The
development of observed structures may be explained by the so-called
''ion-induced grain growth'' as proposed by Hasegawa et al. [11], where it
is thought that surface diffusion causes the adatoms to migrate
preferentially to those grains which are at energetically favored
orientations [12]. Consequently, these grains tend to grow with time and are
transformed eventually to mounds due to minimization of surface free energy.
For thin films, one should consider a further effect of ion beam responsible
for the decay of kinetic roughening at the later stages of bombardment,
namely, the loss of material due to sputter erosion from the sample. After
prolonged bombardment, when the film is thin enough the particle supply
through surface diffusion is supposed to no longer compete with the sputter
removal of atoms as a result of which the mounds cease to grow and
ultimately disappear. For relatively thick metal films, the growth process
continues and the mounds are finally transformed by further erosion to sharp
conical structures [13]. The high $\alpha $ value is an indicative of higher
rate of surface diffusion compared to that of sputtering which is necessary
for the growth and/ or stabilization of the surface structures.\newline

\begin{center}
\smallskip

ACKNOWLEDGMENTS
\end{center}

\noindent \ The authors thank Dr. S. Kundu for the preparation of the Pt
thin films, Mr. A. Das for technical assistance during the AFM measurements
and to IOP, Bhubaneswar for RBS measurements. Finally, the authors are
grateful to Prof. F. Okuyama of Nagoya Institute of Technology for HRSEM
measurements.

\pagebreak

\section*{References}

\begin{itemize}
\item[{[1]}]  {\it Beam Effects, Surface Topography and Depth Profiling in
Surface Analysis}, Eds. A. W. Czanderna, T. E. Madey and C. J. Powell
(Plenum Press, New York, 1998).

\item[{[2]}]  D. Ghose and S. B. Karmohapatro, in: {\it Advances in
Electronics and Electron Physics}, Ed. P. Hawak (Academic Press, New York,
1990), Vol. 79, p. 73.

\item[{[3]}]  S. Facsko, T. Dekorsy, C. Koerdt, C. Trappe, H. Kurz, A. Vogt,
and H. L. Hartnagel, Science {\bf 285}, 1551 (1999).

\item[{[4]}]  A. -L. Barab\'{a}si and H. E. Stanley, {\it Fractal Concepts
in Surface Growth} (Cambridge University Press, Cambridge, England, 1995).

\item[{[5]}]  P. Karmakar, P. Agarwal and D. Ghose, Appl. Surf. Sci. 178
(2001) 83.

\item[{[6]}]  J. J. Vajo, R. E. Doty, and E. -H. Cirlin, J. Vac. Sci.
Technol. A {\bf 14} (1996) 2709.

\item[{[7]}]  F. Family and T. Vicsek, {\it Dynamics of Fractal Surfaces}
(World Scientific, Singapore, 1991).

\item[{[8]}]  R. Cuerno and A. -L. Barab\'{a}si, Phys. Rev. Lett. {\bf 74},
4746 (1995).

\item[{[9]}]  J. T. Drotar, Y. -P. Zhao, T. -M. Lu, and G. -C. Wang, Phys.
Rev. E {\bf 59}, 177 (1999).

\item[{[10]}]  J. H. Jeffries, J. -K. Zuo and M. M. Craig, Phys. Rev. Lett. 
{\bf 76}, 4931 (1996).

\item[{[11]}]  Y. Hasegawa, Y. Fujimoto and F. Okuyama, Surf. Sci. 163
(1985) L781.

\item[{[12]}]  S. Rusponi, G. Costantini, F. Buatier de Mongeot, C. Boragno,
and U. Valbusa, Appl. Phys. Lett. {\bf 75}, 3318 (1999).

\item[{[13]}]  M. Tanemura and F. Okuyama, Nucl. Intrum. Methods B47 (1990)
126.
\end{itemize}

\pagebreak

\section*{Figure captions}

\smallskip

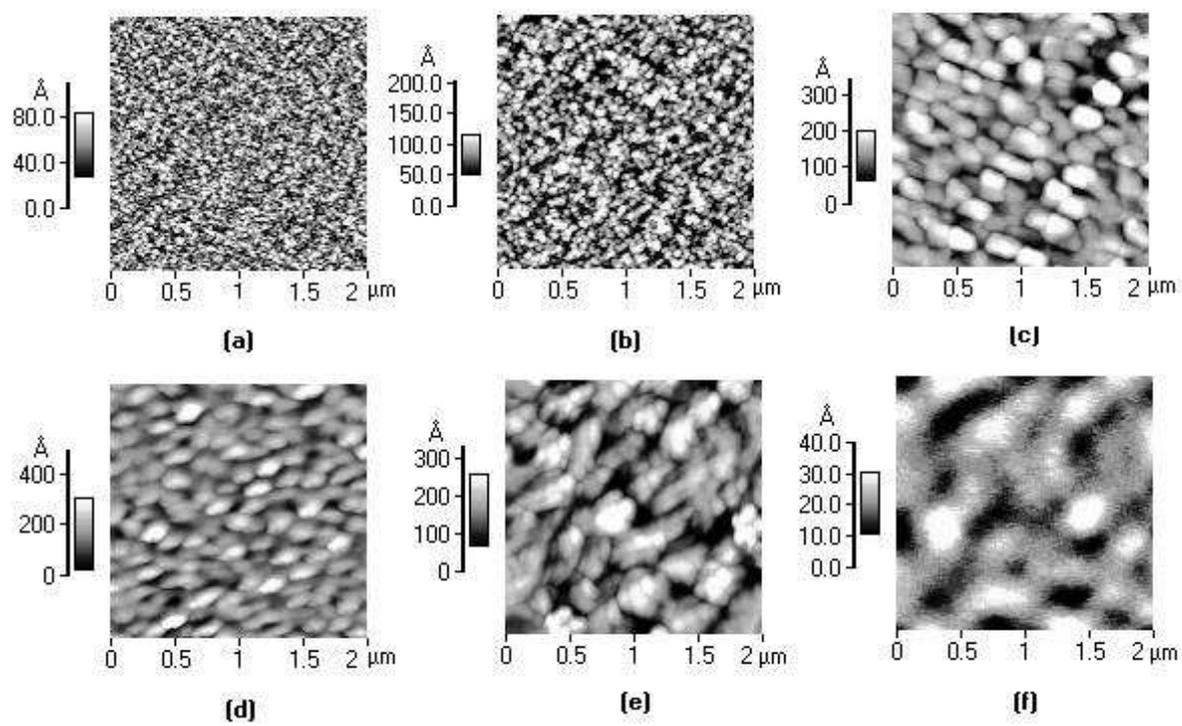
Fig. 1. Some selected AFM images of Ar$^{+}$ sputtered Pt/Si surfaces,
showing a sequence of the evolution of the surface topography with
increasing ion doses: (a) $2\times 10^{15}$ $ions/cm^{2}$; (b) $7\times
10^{15}$ $ions/cm^{2}$; (c) $2\times 10^{16}$ $ions/cm^{2}$; (d) $5\times
10^{16}$ $ions/cm^{2}$; (e) $7\times 10^{16}$ $ions/cm^{2}$ and (f) $2\times
10^{17}$ $ions/cm^{2}$.

\smallskip

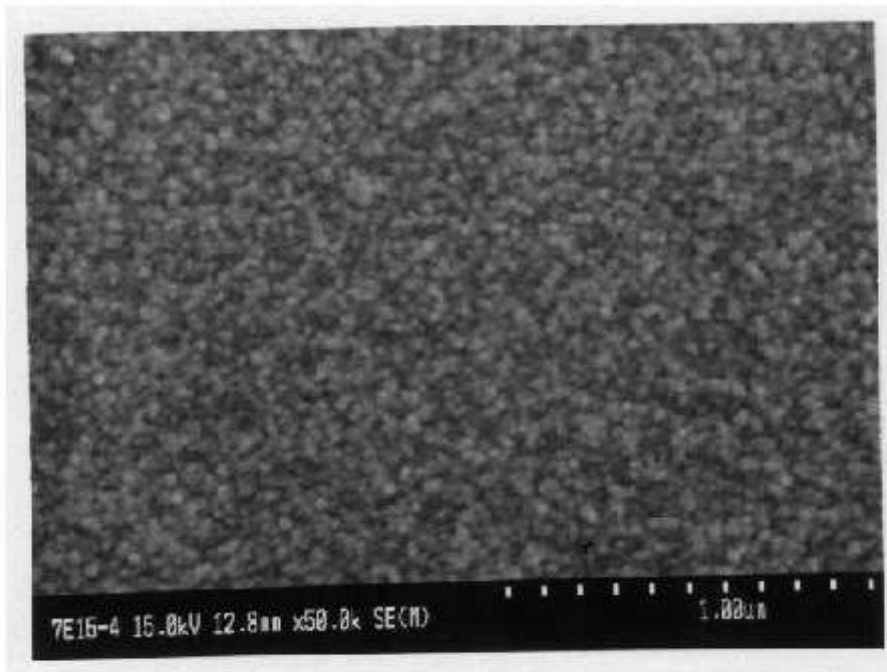
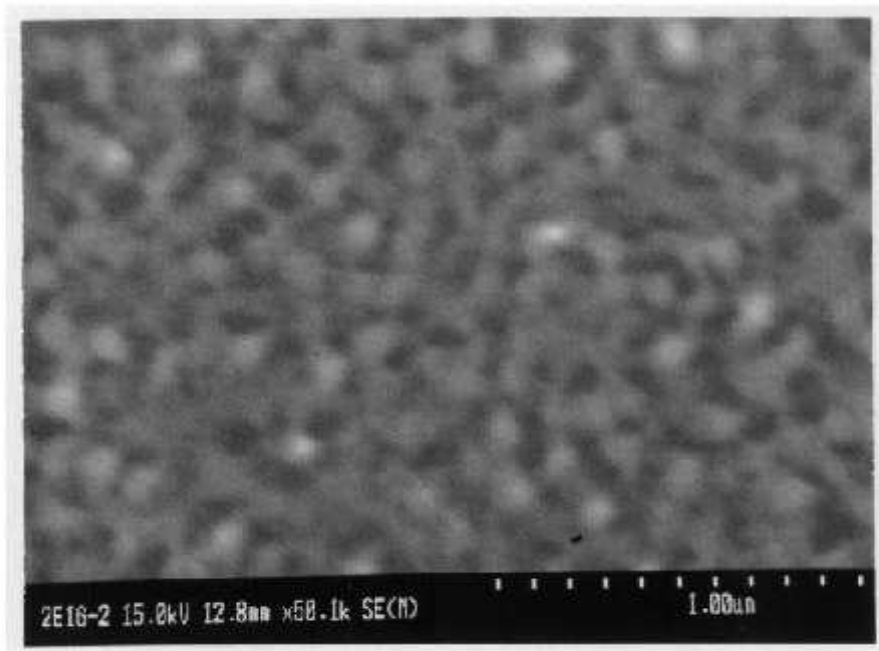
Fig. 2. Typical scanning electron micrographs of Ar$^{+}$ sputtered Pt/Si
surfaces at ion doses (a) $7\times 10^{15}$ $ions/cm^{2}$ and (b) $2\times
10^{16}$ $ions/cm^{2}$.\newline

\smallskip

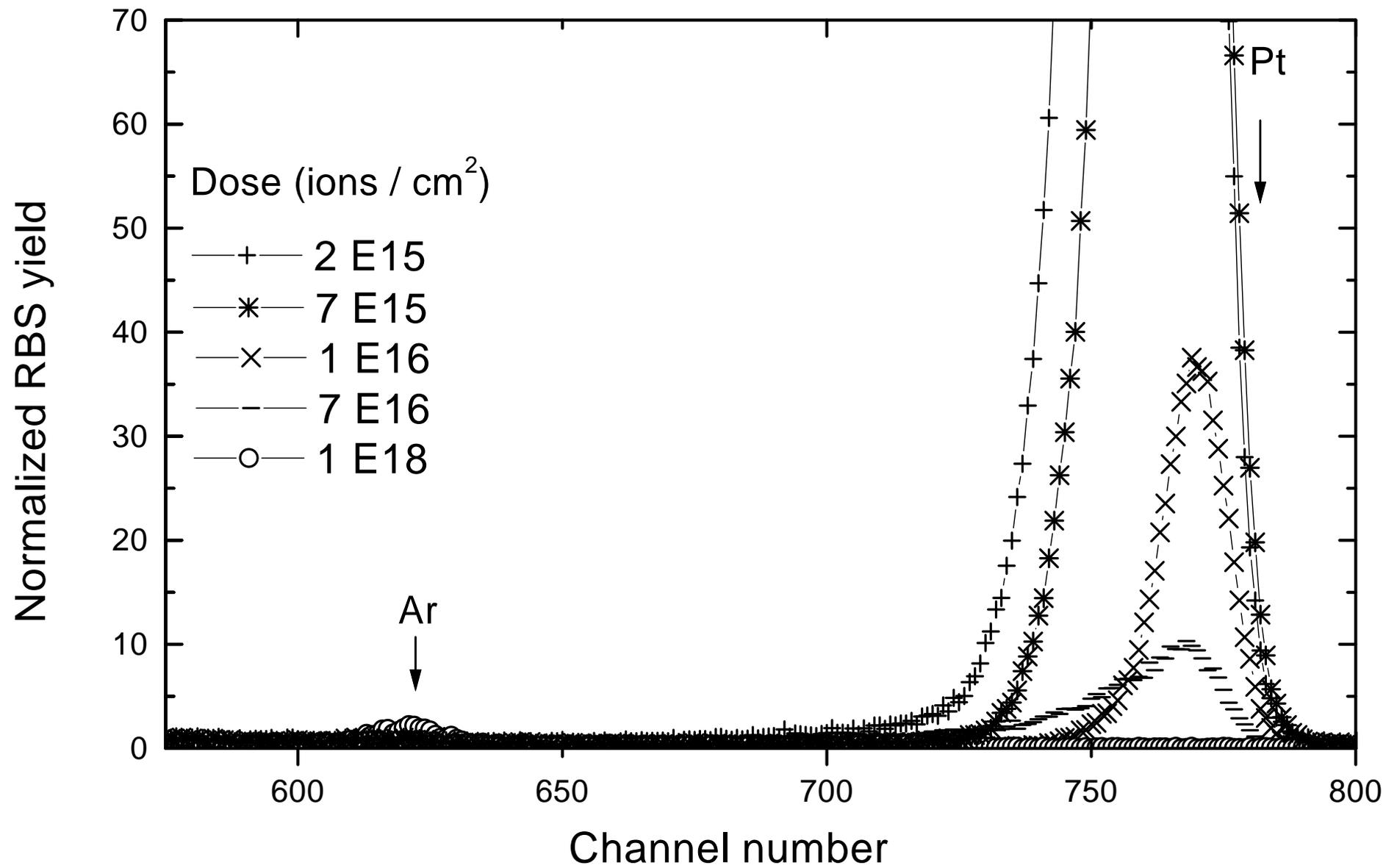
Fig. 3. Pt distribution of RBS spectra for Pt/Si at 8 keV Ar$^{+}$ ion beam
sputtering as a function of ion dose.

\smallskip

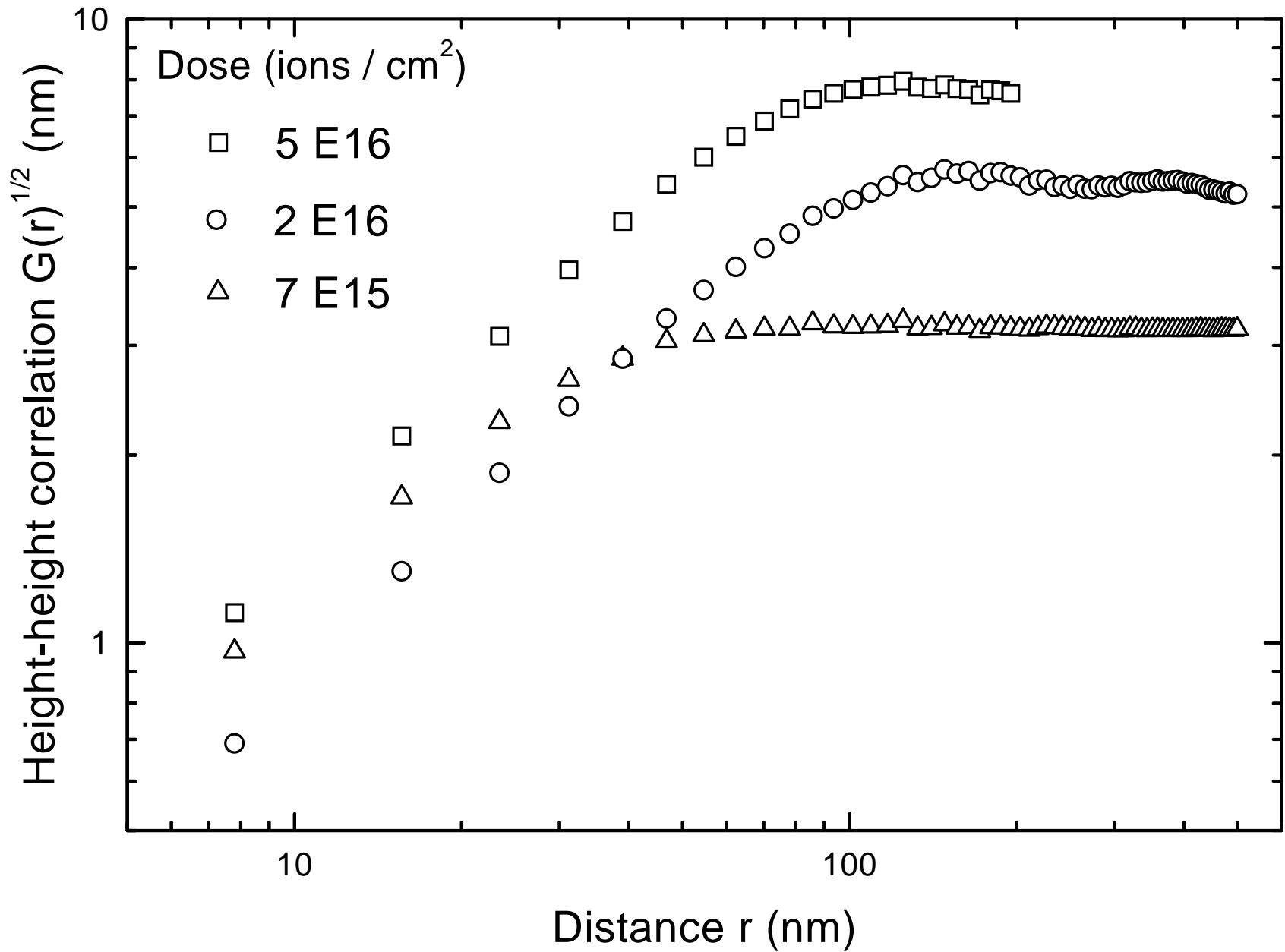
Fig. 4. Some typical log-log plots of the height-height correlation
functions $G(r,$ $t)$, calculated from AFM images, as a function of the
lateral distance $r$ for different bombarding doses $\phi $ as indicated.

\smallskip

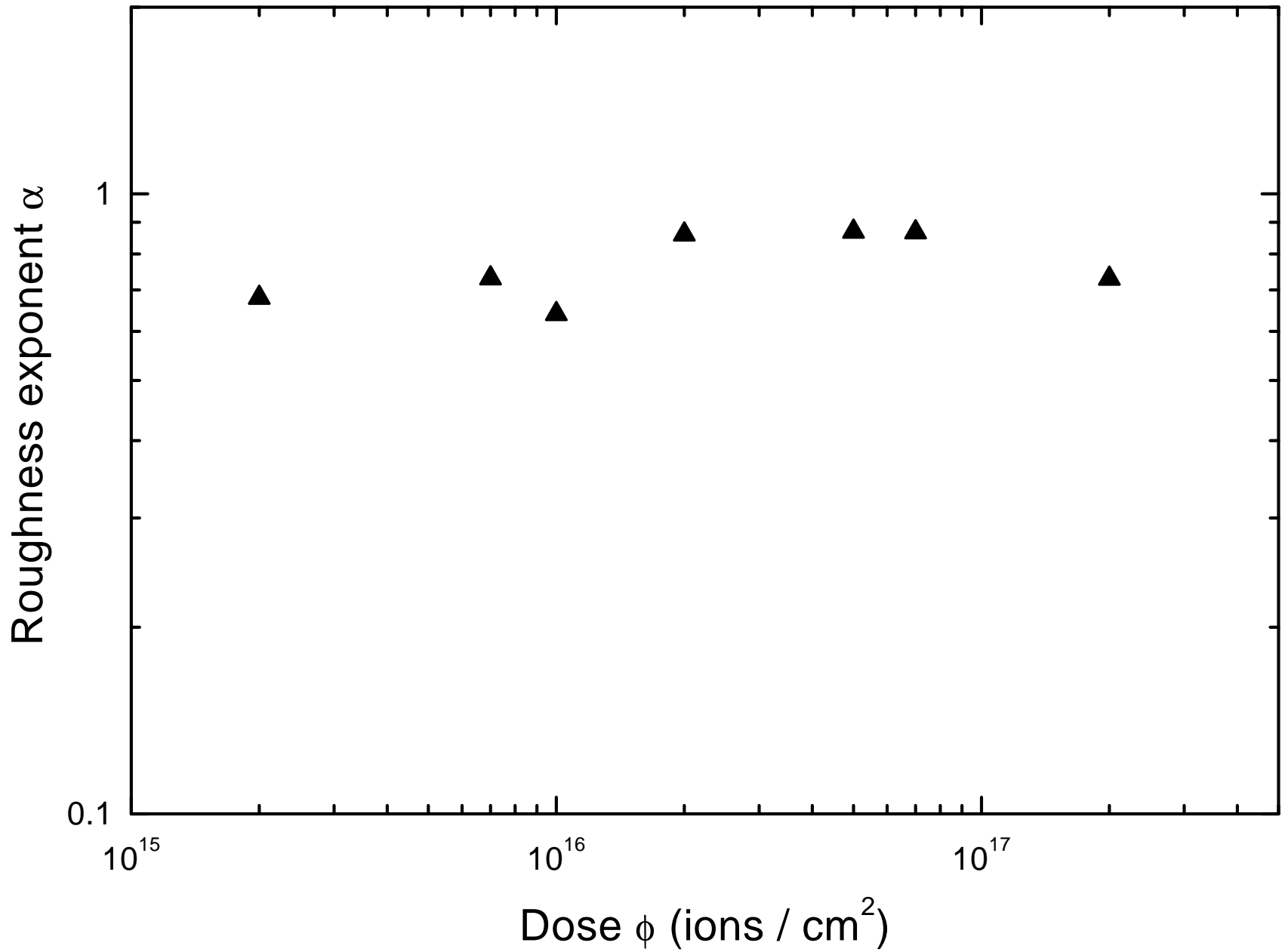
Fig. 5. The plot of the roughness exponent $\alpha $ versus ion dose $\phi $.

\smallskip

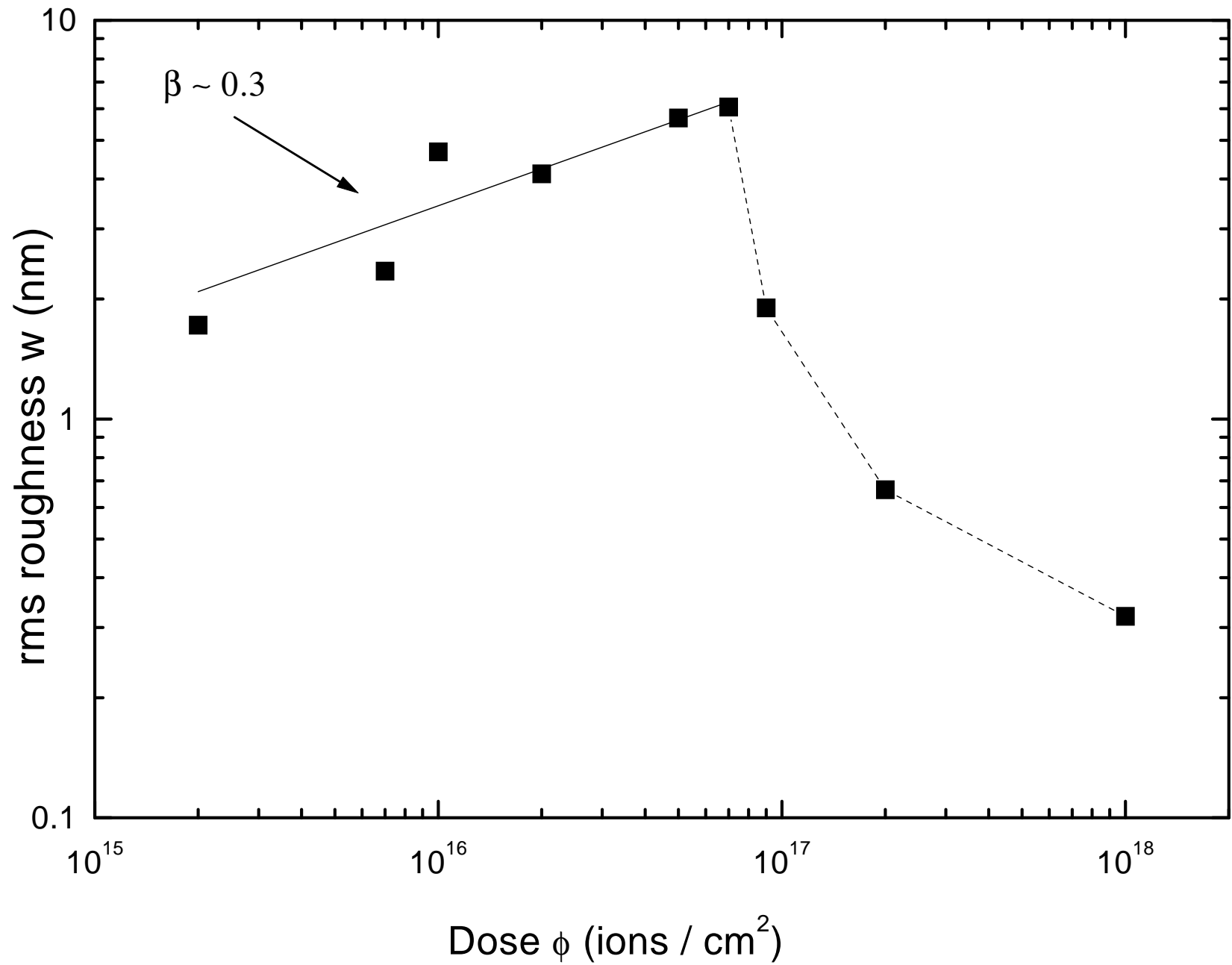
Fig. 6. Showing the rms surface roughness $w$ vs the bombarding ion dose $%
\phi $. The solid line of the growth part represents the fit according to
the power scaling relation $w\sim {\phi }^{\beta }$, while the dashed line
indicates the trend of the data at very high doses.\newline

\smallskip

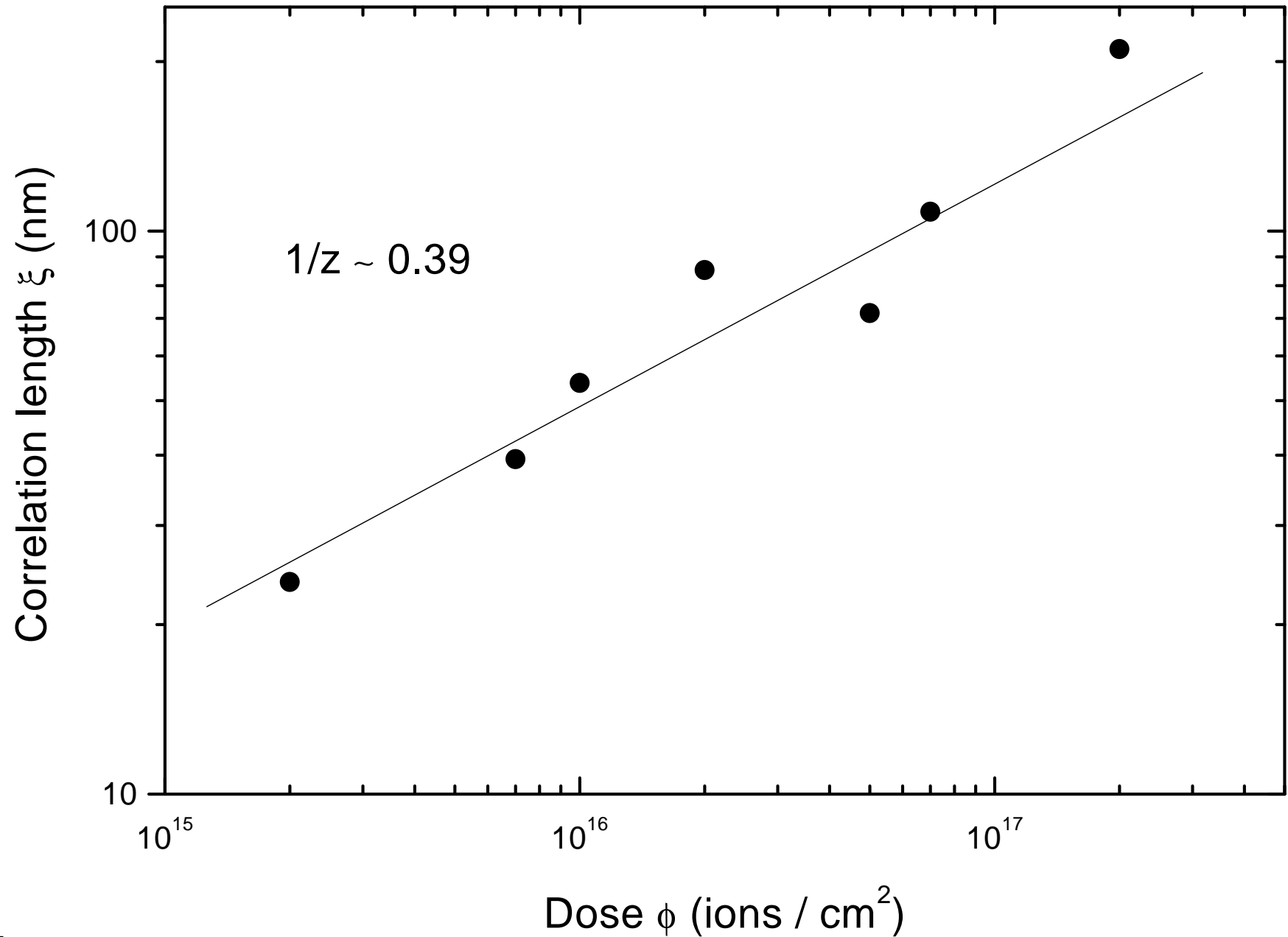
Fig. 7. The plot of the correlation length ${\xi }$ {versus ion dose }$\phi $%
; the fitting line indicates the slope yielding the value of $z$.

\end{document}